# Across-Layer Sliding Ferroelectricity in Graphene-Based Heterolayers: Asymmetry of Next Neighbor Interlayer Couplings

Liu Yang and Menghao Wu

School of Physics and School of Chemistry, Institute of Theoretical Chemistry, Huazhong University of Science and Technology, Wuhan 430074, China

Abstract

Although most two-dimensional (2D) materials are non-ferroelectric with highly symmetric lattices, symmetry breaking may take place in their bilayers upon certain stacking order, giving rise to so-called sliding ferroelectricity where the vertical polarizations can be electrically reversed via interlayer translation. However, it is not supposed to appear in systems like graphene bilayer with centro-symmetry at any stacking configuration, and the origin of the recently reported ferroelectricity in graphene bilayer intercalated between h-BN (Nature 2020, 588, 71) is still unclear. Here we propose a model of across-layer sliding ferroelectricity that arises from the asymmetry of next neighbor interlayer couplings. The vertical polarizations in intercalated centro-symmetric 2D materials like graphene bilayer can be switched via multilayer sliding, and the observed ferroelectric hysteresis can be clarified. Moreover, such ferroelectricity can exist in a series of other heterolayers with quasi-degenerate polar states, like graphene bilayer or trilayer on BN substrate, or even with a molecule layer on surface where each molecule can store 1 bit data independently, resolving the bottleneck issue of sliding ferroelectricity for high-density data storage.

Ferroelectrics are dielectric materials with electrically switchable polarizations, which have been exploited in widespread applications such as memories, radio frequency and microwave devices, etc.. In the past decade, ferroelectricity in atomic-thick 2D materials with van der Waals interface(1) has been increasingly explored for realizing high-density integration of non-volatile random access memories. However, most prevalent two-dimensional materials like graphene and $MoS_2$ with highly symmetrical honeycomb lattices are non-ferroelectric. In 2017 we proposed that for most non-ferroelectric 2D materials, certain stacking of bilayer or multilayer may break the symmetry and give rise to so-called sliding ferroelectricity, where the vertical polarizations are switchable via in-plane translation, i. e., interlayer sliding.(2) Over the next few years, such ferroelectricity have been experimentally detected in bilayer or multilayer BN,(3-5) transition-metal dichalcogenides (TMDs) like $WTe_2$,(6-8) $MoS_2$,(9-14) $ReS_2$,(15) and even multiwall nanotubes(16) and amphidynamic crystal(17). This mechanism is applicable to most 2D materials except for mono-element systems like graphene and phosphorene, as the inversion symmetry of their bilayers can always be maintained at any interlayer-sliding vector. However, ferroelectric-like switching behaviors in bilayer graphene intercalated between h-BN were reported in a recent experiment(18), which cannot be described by the previous model of sliding ferroelectricity. Although the authors attributed the phenomenon to correlation-driven charge transfer that can clarify the anomalous electron transport behaviors, the pure interlayer electron transfer model without ion displacements cannot well explain the observed robust ferroelectric hysteresis as electron can easily go to the ground state by quantum tunneling through barriers. The key elements of a typical ferroelectric model are still missing: what is the structures of its bi-stable polar states, and what is the switching pathway between them?

In this paper we focus on ion displacements and propose a model of so-called across-layer sliding ferroelectricity (ALSF) that stems from the asymmetry of next neighbor interlayer couplings. For the above-mentioned system where bilayer graphene is intercalated between h-BN, such asymmetry gives rise to a considerable vertical polarization that can be reversed via interlayer translation, providing the switching mode to clarify the origin of the reported ferroelectric hysteresis. Such ALSF can also exist in some other multilayer graphene/BN systems with quasi-degenerate polar states, or even with a molecule layer on surface. Herein each molecule can store 1 bit data independently so the data storage density can be enhanced to unprecedented values. It may resolve the bottleneck issue of current sliding ferroelectric homobilayers, with limited density for data storage as large area simultaneous sliding of a whole layer is required during ferroelectric switching(19).

Results and Discussion

The lattice of bilayer graphene possesses inversion symmetry $\hat{I}$ but no horizontal mirror symmetry $\hat{M}_{xy}$. When it is intercalated between h-BN, the two interfacial BN layers attached to the bilayer graphene respectively upwards and downwards can be either parallel or anti-parallel. For the anti-parallel configuration, the inversion symmetry $\hat{I}$ is preserved and forbids the formation of vertical polarization, which will be broken if the two interfacial BN layers are parallel. We have compared the energies of different possible stacking configurations, and the ground state of graphene/BN interface turns out to be staggered with C over B atoms. The bi-stable polar states with the parallel alignment of two interfacial BN layers are displayed in Fig. 1(a), which are correlated by $\hat{M}_{xy}$ operator instead of inversion $\hat{I}$. Although the two graphene/BN interfaces (1st-2nd layer interface and 3rd-4th layer interface) are both staggered with C over B atoms, for one of the polar states, the next-neighbor 1st-3rd layer stacking configuration (AA stacking) is distinct from the 2nd-4th stacking configuration (AB stacking), so the in-equivalence of two graphene layers arises, which is the origin of ALSF. It is vice versa for the other polar state (AB stacking for 1st-3rd layer and AA stacking for the 2nd-4th layer), and such two identical states can be swapped via the simultaneous translation of the two upper C/BN layers with respect to the two down C/BN layers so the bilayer graphene are transformed from AB to BA stacking. Our NEB calculations of such sliding pathway shown in Fig. 1(b) reveal a low switching barrier of only 3.1 meV/unitcell.

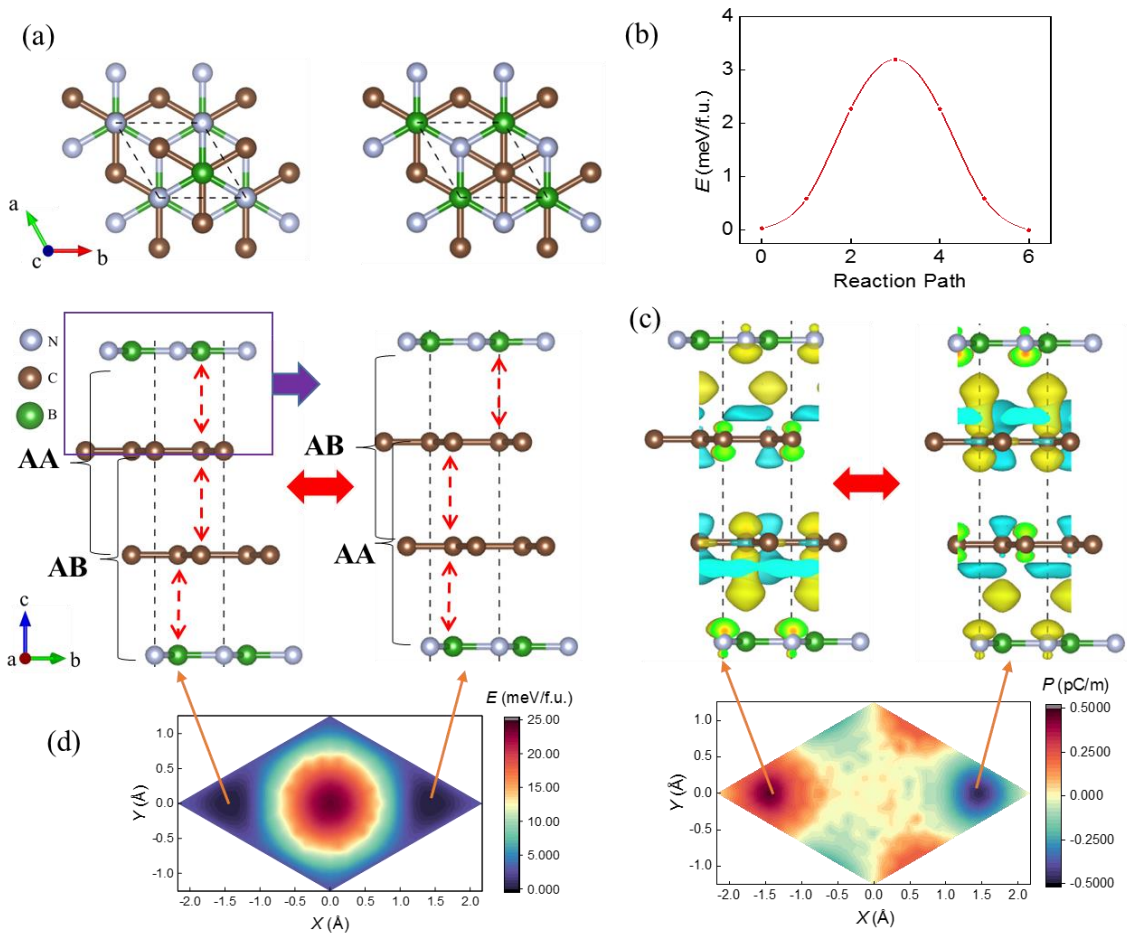

Fig. 1. (a)Geometric structures of the bi-stable polar states for bilayer graphene covered by two parallel BN monolayers, (b)the corresponding switching pathway and (c)differential charge density distributions where yellow and blue isosurfaces respectively indicate electron accumulation and depletion after layer stacking. (d)Energy and polarization landscape of xy plane upon the combined translation of two layers, where (-1.44, 0) and (1.44, 0) correspond to the bi-stable polar states.

The vertical polarization induced by such in-equivalence is estimated to be 0.48 pC/m, smaller than BN bilayer and higher than InSe bilayer.(2) The breaking of symmetry can also be reflected by our Hirshfeld charge analysis, where the charge on two graphene layers are respectively 0.0121 and 0.0105e per unitcell, which are both around -0.0115e for the two interfacial BN layers.

The electron distribution in Fig. 1(c) also shows the evidence of such interlayer in-equivalence, where the distribution of π electron cloud of the C atoms of two layers are greatly differentiated by their different across-layer configurations. The 2D landscape for energy and polarization is plotted in Fig. 1(d), where x and y denote the translation between the upper graphene layer and the down graphene layer, and the two minima (-1.44, 0) and (1.44, 0) in the energy landscape corresponding to AB and BA stacking of two polar states with opposite polarizations can be visualized.

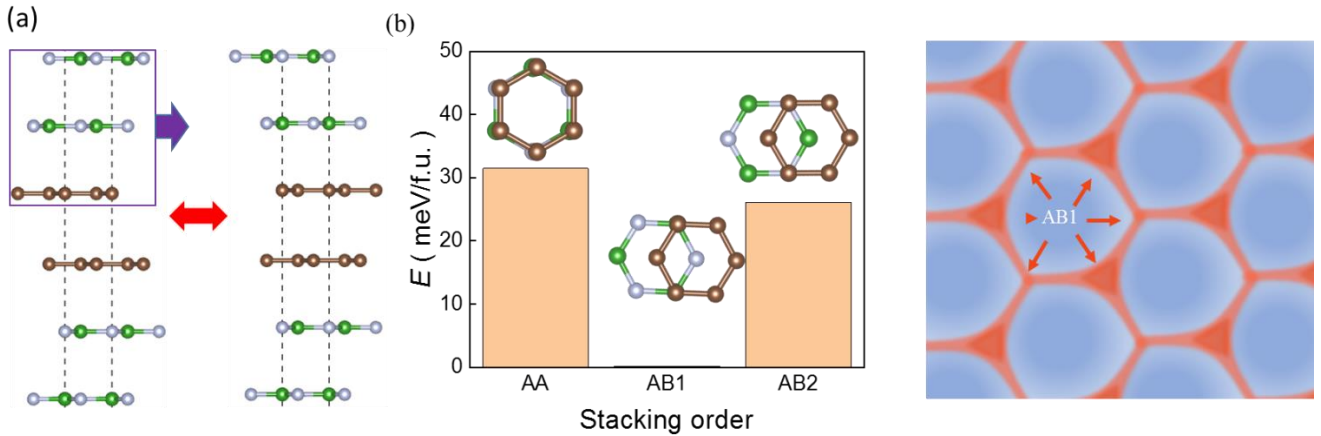

Fig. 2. (a)Geometric structure of the bi-stable polar states for bilayer graphene encapsulated by two BN bilayers, (b) Comparison of relative energy for different stacking orders of graphene/BN hetero-bilayer, where the AB1 domains will still occupy most of the region.

If the two non-centrosymmetric BN monolayer in Fig. 1 are replaced by two nonpolar BN bilayers, as shown in Fig. 2(a), similar ALSF can still emerge with slightly enhanced polarization of 0.51 pC/m. It seems that such ALSF model is applicable to bilayer graphene intercalated in h-BN of any thickness. However, due to lattice mismatch, herein Moire pattern is likely to locate at graphene/BN interface, e.g., the H4 device in Ref. (18). It has been reported that graphene can either stretch to adapt to a slightly different hBN periodicity in a commensurate state or exhibit little adjustment in the incommensurate state, and in the former case, areas with matching lattice constants are separated by domain walls that accumulate the generated strain(20). For the latter case, according to previous studies on similar Moire patterns,(21, 22) the domains of the lowest-energy stacking configuration will be maximized by reconstruction. Our calculations turn out that the AB1 with C over B is much lower in energy compared with other configurations, so the reconstruction will give rise to discrete commensurate AB1 domains divided by narrow domain walls (see Fig. 2(b)), which accords with previous experimental observations on twisted graphene/BN hetero-bilayer.(23) Therefore in both cases, AB1 domains will still dominate and ALSF switchable via simultaneous sliding of the upper two layers in Fig. 1 will maintain. The authors of Ref. (18) have explained the anomalous electron transport behaviors using the model of correlation-driven charge transfer induced by Moire minibands, while our ALSF model on the ion lattice provides the switching barrier between bi-stable polar states for ferroelectric hysteresis, and their combination may give the full picture for H4 device.

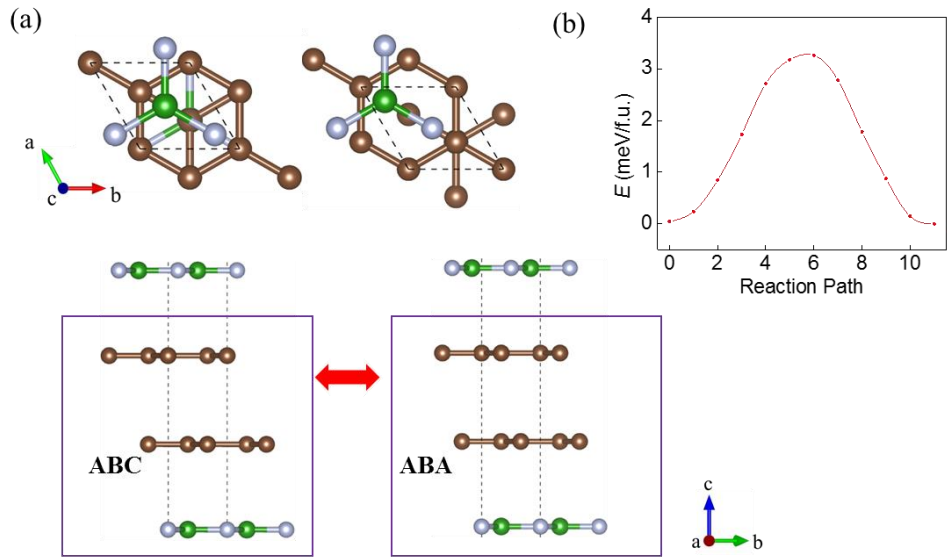

Fig. 3. (a)Geometric structures of the two quasi-degenerate polar states for bilayer graphene covered by two BN monolayers, with a twist angle of 30 degree between the upper BN monolayer and its adjacent graphene layer, (b)the switching pathway between two states.

We also try to shed light on the origin of previously reported ferroelectricity in another device H2 (18, 24) where bilayer graphene was intercalated in BN with a relative angle around 30 degree between the top interfacial BN layer and graphene bilayer. As mentioned above, the down C/BN interface without twisted angle is inclined to be staggered with C over B where C-B distance is much shorter compared with C-N distance, and the interlayer potential will change with the varying of such difference via interlayer sliding(2). In contrast, for the upper C/BN interface with a twist angle of 30 degree that give rise to 2D quasi-crystal,(25) the two layers can be deemed as decoupled and the average interlayer C-B and C-N distances are identical, which scarcely depend on interlayer translation. Although it is challenging to simulate the quasi-crystal of graphene/BN with a twist angle of 30 degree in periodic model, the structure can be equivalent to the configuration with the same interlayer C-B and C-N distances for the upper graphene/BN (1st-2nd) interface in Fig. 3(a). The stacking of the graphene bilayer with the down BN layer can be either ABA or ABC, with the same 2nd-3rd and 3rd-4th adjacent layer stacking configurations but distinct across-layer 2nd-4th stacking configurations. The two states are not exactly identical but almost degenerate in energy (with a small difference of only 0.05 meV/unitcell), much lower compared with the switching barrier via interlayer translation (over 3 meV/unitcell according to the NEB calculation of pathway in Fig. 3(b)). Meanwhile the different next-neighbor 2nd-4th interlayer couplings of two states (AA and AC) gives rise to distinct vertical polarizations (respectively 1.70 pC/m and 0.76 pC/m), indicating the existence of ALSF with a switchable polarization of 0.47 pC/m.

If the top BN layer is removed, as shown in Fig. 4(a), the two lowest-energy configurations for bilayer graphene on BN monolayer are still ABA and ABC state, with the same adjacent layer stacking and different across-layer stacking. Their energy difference will increase to 0.31 meV/unitcell, still much lower compared with the sliding barrier over 3 meV/unitcell. Hence the two states can still be deemed as quasi-degenerate, and their difference in vertical polarizations (respectively 2.29 pC/m and 1.81 pC/m) indicates a switchable polarization of 0.24 pC/m. Similarly, if the top graphene layer is further substituted by a layer of benzene molecules (see Fig. 4(b)), where each benzene will be inclined to form AB stacking with the attached graphene layer, the energy difference between ABA and ABC state will be 0.32 meV/unitcell, and the difference of two states in vertical polarizations indicates a switchable polarization of 0.035 pC/m. As summarized in the previous review of sliding ferroelectricity for homobilayers(19), due to the in-plane rigidity of 2D layer, the density for data storage is limited since large-area simultaneous sliding of a whole layer is required during ferroelectric switching. Herein this issue can be resolved where each benzene molecule may store 1 bit data independently, giving rise to ultrahigh areal density up to $10^4$ Tbit/inch$^2$(orders of magnitude higher compared with the superparamagnetic limit of hard disk around 40 Gbit/inch$^2$).[26] It can also be substituted by larger benzene-based flake or molecule (e.g., triphenylene, coronene, which will also form into AB stacking with the attached graphene layer) for higher switchable polarizations and barriers. The quasi-degenerate polar states also exist in trilayer graphene on BN monolayer, as shown in Fig. 5 where the ground state ACBA is only 0. 28 meV/unitcell lower in energy compared with CABA state, while other states like BABA and BCBA are over 2 meV/unitcell higher in energy. For the ACBA and CABA states with the same adjacent layer stacking and different across-layer stacking, their difference in vertical polarizations implies a switchable polarization of 0.30 pC/m.

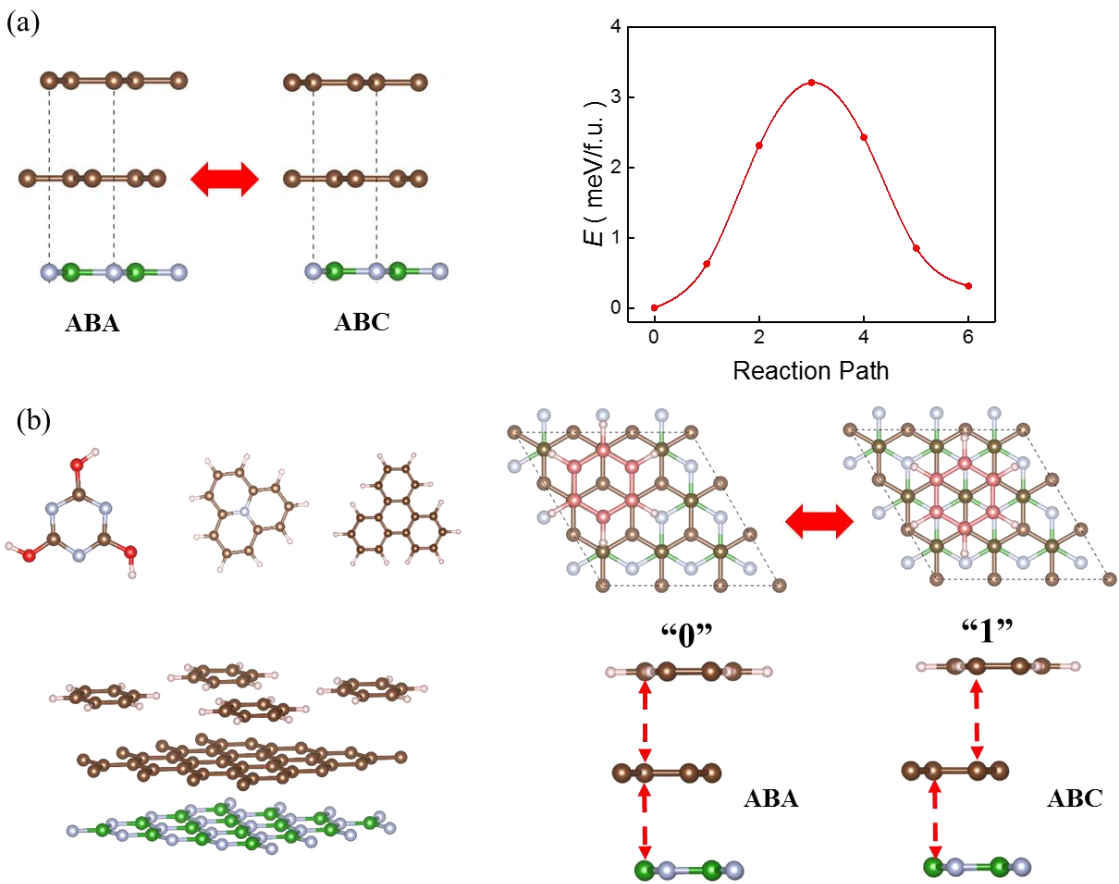

Fig. 4. (a) The switching pathway between two quasi-degenerate ABA and ABC polar states for graphene bilayer/BN monolayer. (b) As the top graphene layer is substituted by a layer of benzene-based molecules forming into AB stacking with the attached graphene below, the two quasi-degenerate ABA and ABC polar states will enable independent single-molecule data storage.

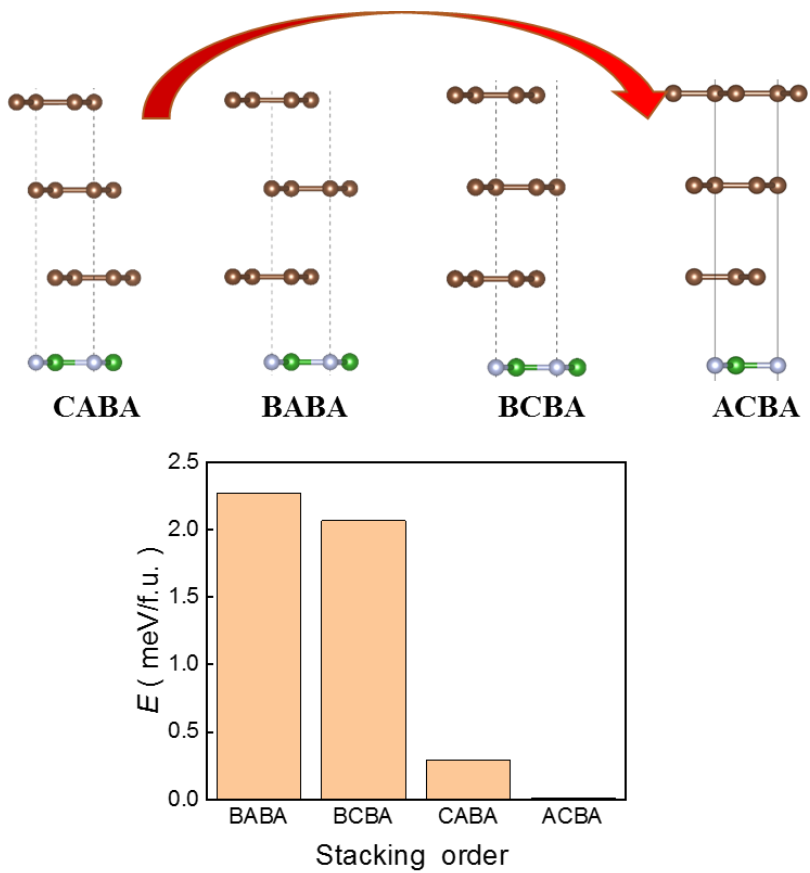

Fig. 5. Four stacking configurations of graphene-trilayer/BN-monolayer, where CABA and ACBA possess two lowest energies with distinct vertical polarizations.

In conclusion, we have proposed the model of across-layer sliding ferroelectricity with asymmetry arising from the next neighbor interlayer couplings and the induced vertical polarization can be reversed via interlayer translation, unveiling the ion displacement origin of previously observed ferroelectric hysteresis of graphene bilayer intercalated in BN. Similar mechanism can be applied to some other multilayer graphene/BN systems with quasi-degenerate polar states. In particular, such ferroelectricity can be used for data storage with unprecedented areal density up to $10^4$ Tbit/inch$^2$ in graphene/BN system with a benzene-based molecule layer on the surface.

Methods

Theoretical calculations involved in this paper were performed by using density functional theory (DFT) implemented in the Vienna ab initio Simulation Package (VASP 5.4) code(27, 28). The projector augmented wave (PAW) potentials(29) with the generalized gradient approximation (GGA) in the Perdew–Burke–Ernzerhof (PBE)(30) form were used to treat the electron-ion interactions. The DFT-D2 functional of Grimme (31) was used to describe the van der Waals interactions. A large vacuum region with a thickness of 25 Å was added in the z direction to minish interaction between adjacent slabs. The first Brillouin zone (BZ) was sampled with Monkhorst–Pack meshes method(32) at the Γ center. The 15×15×1 k-point grid was implemented for graphene/BN systems. The kinetic energy cutoff is set to be 520 eV and energy convergence criterion was set to $10^{-5}$ eV. The force convergence was set to 0.01 eV/Å for the geometry optimization. Dipole correction(33) and the Berry-phase method(34) were employed to evaluate the vertical polarization. The ferroelectric switching pathways was obtained by using nudged elastic band (NEB) method(35).

**Acknowledgements**

This work is supported by the National Natural Science Foundation of China (Nos. 22073034).

**Author contributions:** M. W. designed and supervised research; L. Y. performed research; L. Y. and M. W. wrote the paper.

**Competing interests:** The authors declare no competing interests.